# Towards a consensus on rheological models for elastography in soft tissues

K J Parker[1], T Szabo[2], S Holm[3],

[1] Department of Electrical and Computer Engineering, University of Rochester, 724 Computer Studies Building, Box 270231, Rochester, NY 14627, USA

[2] Department of Biomedical Engineering, Boston University, 44 Cummington Mall, Room 403, Boston, MA 02215, USA

[3] Department of Informatics, University of Oslo, Postboks 1080, Blindern, 0316 Oslo, Norway

Email:  kevin.parker@rochester.edu

**Abstract**
A rising wave of technologies and instruments are enabling more labs and clinics to make a variety of measurements related to tissue viscoelastic properties.  These instruments include elastography imaging scanners, rheological shear viscometers, and a variety of calibrated stress-strain analyzers. From these many sources of disparate data, a common step in analyzing results is to fit the measurements of tissue response to some viscoelastic model.  In the best scenario, this places the measurements within a theoretical framework and enables meaningful comparisons of the parameters against other types of tissues.  However, there is a large set of established rheological models, even within the class of linear, causal, viscoelastic solid models, so which of these should be chosen?  Is it simply a matter of best fit to a minimum mean squared error of the model to several data points?  We argue that the long history of biomechanics, including the concept of the extended relaxation spectrum, along with data collected from viscoelastic soft tissues over an extended range of times and frequencies, and the theoretical framework of multiple relaxation models which model the multi-scale nature of physical tissues, all lead to the conclusion that fractional derivative models represent the most succinct and meaningful models of soft tissue viscoelastic behavior.  These arguments are presented with the goal of clarifying some distinctions between, and consequences of, some of the most commonly used models, and with the longer term goal of reaching a consensus among different sub-fields in acoustics, biomechanics, and elastography that have common interests in comparing tissue measurements.

# 1. Introduction

There has been a robust proliferation of techniques that can estimate tissue stiffness and viscoelastic properties using advanced imaging techniques (Parker *et al.*, 2011; Li and Cao, 2017; Zvietcovich *et al.*, 2017). Many of these techniques employ shear waves in tissue, in the frequency range of 10 – 1000 Hz and so the abundance of new data from shear wave propagation forces a re-examination of rheological models of the viscoelasticity of tissues. Put another way, many investigators now have measurement tools to track shear waves in tissues, and would like to incorporate the simplest, most accurate, and meaningful viscoelastic model available that can capture the essential time domain and frequency domain characteristics of the behaviors they observe, while extracting the most meaningful parameters for diagnostic value. However, which model should one choose given the large range of available choices that have been developed over a century of biomechanics and related fields (Fung, 1981a; Lakes, 1999a; Bilston, 2018; Nyborg, 1975)?

While different models have been compared in the past (Klatt *et al.*, 2007; Catheline *et al.*, 2004; Sinkus *et al.*, 2007; Urban *et al.*, 2011; Zhou and Zhang, 2018), our study takes into account a larger set of advanced viscoelastic models as well as a growing body of viscoelastic studies of the liver to build a consistent rationale for model selection. In the case of determining the shear modulus, most commonly applied measurement methodologies may provide only a handful of points through which many models can be fit. Furthermore, even though the shear wave speed or modulus is known to vary strongly with frequency, often only one value is reported. Which approach is best?

In addressing these common problems, the three authors as independent researchers have reached similar conclusions and offer a synoptic perspective as a first step towards a possible larger consensus on appropriate viscoelastic models for shear wave elastography in soft tissues. To argue this, we consider experimental data from tissues over a range of conditions. Then we examine some of the simplest (but inappropriate) models, working up in complexity and then generalizing over multiple scales to a very useful simplification. Fractional derivative models and consequent power law behavior are seen to possess a desirable combination of simplicity and utility and are recommended for general use.

It is necessary to define the terms and focus of this work. By "soft tissues", we mean macroscopically homogeneous and isotropic normal tissues such as the liver, prostate, thyroid, and possibly others such as the brain. We consider only small strain, linear models and exclude guided waves in structures. This excludes from consideration muscle, tendon, arterial walls, cornea, bone, and large strain conditions. These require additional considerations and are beyond the scope of this work.

Nonetheless, within the scope of soft tissues lie some of the major successes of elastographic imaging techniques, notably the quantification of liver stiffness relevant to staging of fibrosis (Cosgrove *et al.*, 2013; Barr *et al.*, 2015). To further improve the role of elastography, a multicenter study, QIBA, the Quantitative Imaging Biomarker Alliance, is examining the consistency of different approaches to clinical measurements of the shear modulus with ultrasound (Palmeri *et al.*, 2015; RSNA/QIBA, 2012; Hall *et al.*, 2013). Agreement on appropriateness of models for these soft tissues would be a significant first step in creating a common and coherent framework for comparing results from a variety of sources. Accordingly, this paper is organized to make a coherent argument starting with "ground truth" from experimental measures of extended time domain and frequency domain viscoelastic behaviors of tissue. Then we consider some popular simple models, and the pathway towards more realistic rheological models for soft tissues.

Furthermore, since the viscoelastic nature of tissues implies that the real and imaginary elastic moduli and therefore absorption and shear wave speed are strong functions of frequency, the relationships between these four parameters are reviewed in **Appendix 1**. However, different notations add to the confusion and different interpretations about model applicability. Because these differences in notations from allied branches of science have diverged, we introduce a glossary of terms in **Appendix 2**. Issues with fitting data to different models include the usual case in which only a few points are measured over a narrow frequency range. To address these challenges within a coherent framework, we begin with viscoelastic tissue measurements taken by different methodologies over a wide range of frequencies.

## 2. Tissue behavior from published results

*2.1 Selection criteria*

There are a great many estimates of soft tissue biomechanical properties, covering a wide range of measurement techniques, sample conditions, and results. Elastography techniques that utilize shear waves commonly excite the waves by external sources applied to the surface, or by acoustic radiation force push pulses within the organ of interest. Our focus here is on characterizing the tissue properties that govern the tissue displacements and shear wave propagation from any source. For our purposes, we have sought out measurements that satisfy the following conditions:

- The published estimates are crossed checked by independent measures.
- The estimates cover a decade or more in time or frequency.
- The estimates are obtained for more than one sample.

- The data are presented in graphic or tabular form.
- We also assume that the data have been corrected for any extra artifacts of the shear generation and detection processes.

We included data from several measurement methods including dynamic mechanical testing, resonance, magnetic resonance elastography (MRE), and shear wave elastography (SWE) which involve wave propagation.

*2.2 Published data*

In this section, we illustrate the wide range of data types in the literature including time and frequency data sets as well as those utilizing rheological parameters and shear wave velocities. Some common parameters and their inter-relationships are given in **Appendices 1 and 2**.

One early reference on liver properties is Liu and Bilston (2000) who employed a number of measurements of linear, small strain (less than 1%) conditions. Their stress relaxation results, (**see Figure 1**) demonstrate a reasonably straight line on log-log scale over many decades of duration during the hold. The measured stress suggests a $1/t^a$ power law behavior.

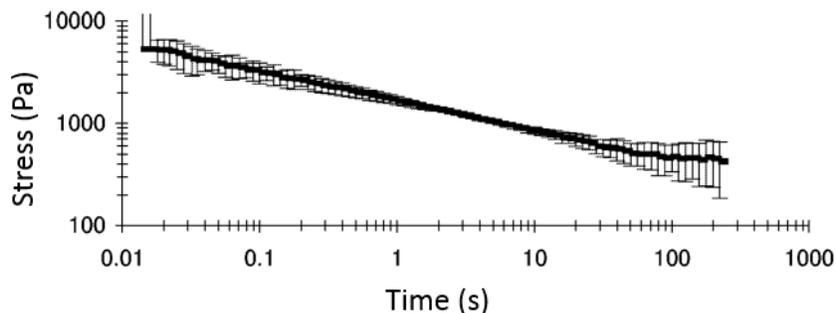

**Figure 1. Stress relaxation results from two liver samples. Vertical axis: measured stress, horizontal axis time in seconds after step compression, on a log-log scale. The behavior is consistent with a power law relaxation. Reprinted from *Biorheology*, vol. 37, Liu and Bilston, "On the viscoelastic character of liver tissue: experiments and modelling of the linear behavior," pp. 191-201 (2000).**

Another example, but measuring shear wave speed as a function of frequency, comes from Ormachea *et al.* (2016), where three different measurements were assessed for their compatibility: crawling waves (generated from external sources), single line tracking shear waves (generated from push pulses), and stress relaxation (time domain results curve fit through the Kelvin-Voigt fractional derivative (KVFD) model to frequency domain predictions). Their results are shown in **Figure 2** on a log-log scale. Their general trend is also consistent with power law behavior over the shear wave propagation range of 40 – 400 Hz.

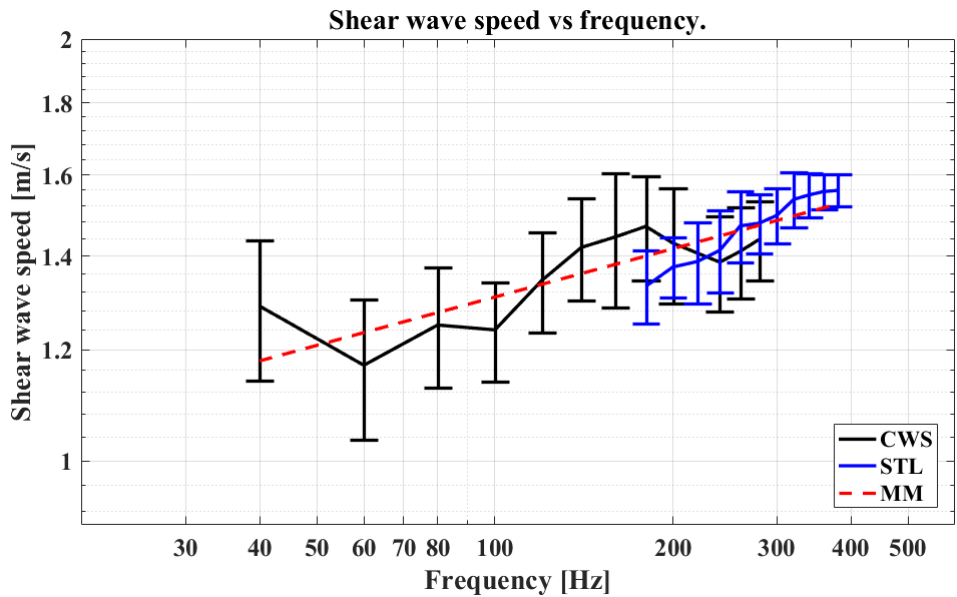

**Figure 2. Shear wave speed $c_s(f)$ data extracted from Ormachea *et al*. (2016) from bovine liver samples, including estimates derived from crawling waves (CWS), single tracking line shear wave estimators (STL), and curve-fit based on stress relaxation results (MM). The nearly linear (on log-log scale) combined results are consistent with the concept of power law behavior.**

A third example demonstrates estimates of shear modulus as a function of frequency, from viscometry and MRE results in bovine and human liver from Klatt *et al*. (2007; 2010). Their overall combined results are consistent with power law behavior between 2 and 70 Hz (see **Figure 3).**

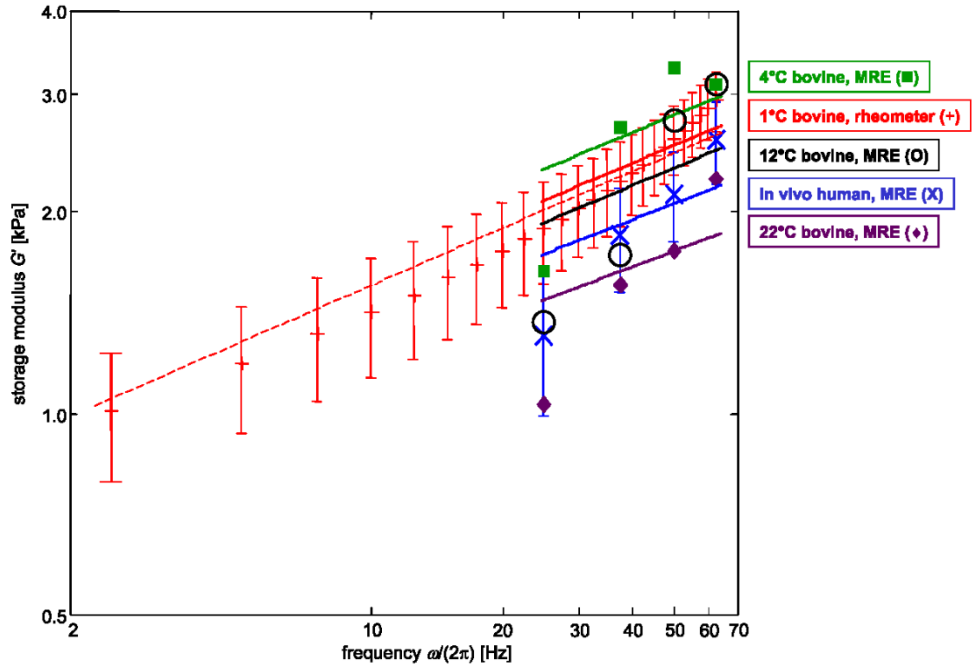

**Figure 3. Storage modulus $E'(\omega)$: The experimental data and the fitted functions according to the spring-pot model are represented by symbols and lines, respectively. The data obtained by the rheometer tests and the *in vivo* examinations represent mean values. The error bars correspond to the standard deviation. Rheometer data were separately fitted once within the entire displayed frequency range and secondly within the dynamic range of multifrequency MRE. Reprinted from *Biorheology*, vol. 47, Klatt *et al*., "Viscoelastic properties of liver measured by oscillatory rheometry and multifrequency magnetic resonance elastography," pp. 133-141 (2010).**

Another example from Kiss *et al*. (2004) is shown in **Figure 4**. These investigators measured the complex Young's modulus of groups of fresh bovine liver samples with and without thermal lesions using a rheometer. The instrument was considered to be accurate below 100 Hz, the rightmost data points represent likely artifacts. These data were fit to a fractional derivative model (to be described in section 3, solid lines) with reasonable accuracy.

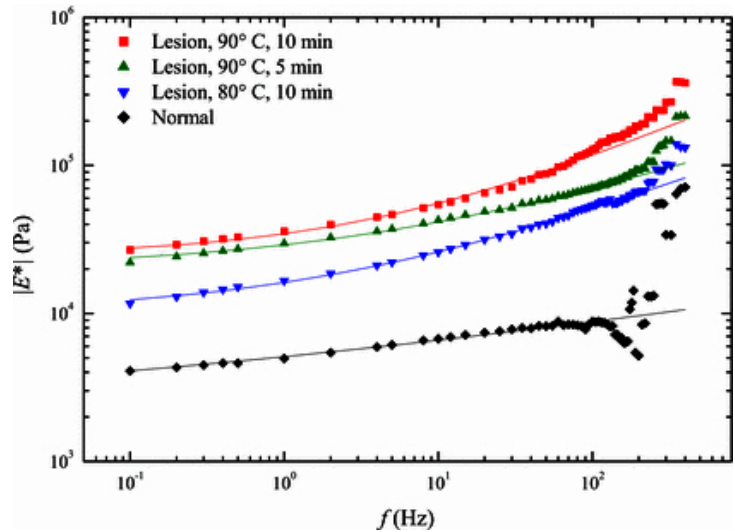

**Figure 4. Magnitude of Young's modulus** $E_0(f) = 3\mu(f)$ **in fresh bovine samples and those altered by diathermy. Data from a rheological instrument are shown in symbols, and theoretical fits to a fractional derivative model is shown as solid lines. Reprinted from** *Physics in Medicine and Biology*, **vol. 49, Kiss** *et al.*, **"Viscoelastic characterization of** *in vitro* **canine tissue," pp. 4207-4218 (2004).**

Finally, we draw a comprehensive perspective of broadband data for liver by drawing from five different sources and measurement methodologies. **Figure 5** combines soft tissue data from Bilston (2018), Kiss *et al*. (2004), Klatt *et al*. (2010), Wex *et al*. (2013), and Chatelin *et al*. (2011) over four decades of frequency, and demonstrates that the shear elastic modulus follows a generally increasing power law across many decades of frequency. These power law relationships indicate structure extending across several decades of measurement scales.

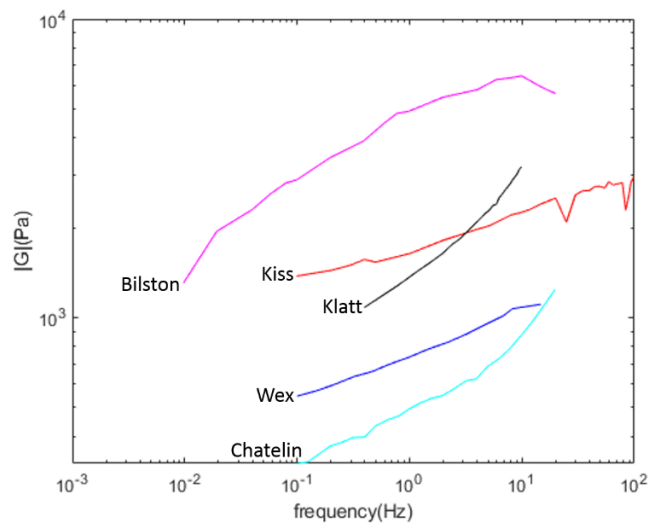

**Figure 5.** Compilation of soft tissue data from different publications (see text for specific references). Vertical axis, estimates of shear modulus in Pa for liver tissues. Horizontal axis, frequency of measurement. The general trend shows increasing values on a log-log plot extending over four decades of frequency; a power law would be represented as a straight line in log-log space.

## 3. Models of tissue response

In this section we review first the simplest and well known classical viscoelastic models. After considering their behavior the discussion moves on to more general and, we argue, more realistic models that can capture power law phenomena with only 2, 3, or 4 parameters depending on the degree of approximation that circumstances allow for.

*3.1 Definitions*

Because tissues are viscoelastic, simple elastic relationships between shear wave speed and elastic moduli, such as $c_s = \sqrt{\mu/\rho}$ (where $c_s$ is shear wave speed and $\rho$ is density), no longer apply. Instead of responding immediately to a change in applied strain or stress, tissue responds slowly or relaxes; therefore, moduli are functions of time or alternatively, frequency. Similarly a simple Hooke's law in which stress ($\sigma$) is proportional to strain ($\varepsilon$), no longer applies. Here we introduce a modified viscoelastic Hooke's law common in mechanical testing:

$$\sigma_{IR}(t) = \sigma_{SR}(t) * \frac{\partial \varepsilon(t)}{\partial t} = \frac{\partial \sigma_{SR}(t)}{\partial t} * \varepsilon(t), \tag{1}$$

where the stress response to a unit step in shear strain is called the stress relaxation response, $\sigma_{SR}(t)$, and its Fourier transform is $\tilde{\sigma}_{SR}(\omega)$. The Fourier transform of $\sigma_{IR}(t)$ is a complex function $\tilde{\sigma}_{SR}(\omega)$. Likewise, the stress response to an impulse in strain, more common in acoustics, is the elastic response to an impulse $\mu(t) = \sigma_{IR}(t)$, and its Fourier transform, the transfer function, is the dynamic modulus $\tilde{\mu}(\omega) = i\omega \tilde{\sigma}_{SR}(\omega)$. Another example of stress relaxation is shown in

**Figure 1**. The relationship of these moduli to other notations and shear wave propagation parameters can be found in **Appendices 1 and 2**.

*3.2 Classical models*

The classical linear viscoelastic models are the Zener, Kelvin-Voigt, and Maxwell models shown in **Figure 6**. The Zener model is called the standard linear solid model which has a single relaxation process. The Kelvin-Voigt and Maxwell models are both subsets of the Zener model. The responses for these models are the following (Holm, 2019b):

$$\text{Zener:}\quad \frac{\sigma(\omega)}{\varepsilon(\omega)} = \mu_e \left[\frac{1+i\omega\tau_\varepsilon}{1+i\omega\tau_\sigma}\right]$$

$$\text{Kelvin-Voigt:}\quad \frac{\sigma(\omega)}{\varepsilon(\omega)} = \mu_e + i\omega\eta \quad (2)$$

$$\text{Maxwell:}\frac{\sigma(\omega)}{\varepsilon(\omega)} = \left[\frac{i\omega\eta}{\mu_e + i\omega\eta}\right],$$

in which

$$\tau_\sigma = \frac{\eta}{\mu},\ \tau_\varepsilon = \eta\left(\frac{1}{\mu}+\frac{1}{\mu_e}\right) \geq \tau_\sigma. \quad (3)$$

For instance, the Kelvin-Voigt model is a low-frequency approximation to the Zener model where the effect of the spring $\mu$ can be neglected as $|\omega\eta| \ll \mu$. Likewise, the Maxwell model is a fluid model since when the frequency approaches zero its relaxation modulus $(1/\mu + 1/i\omega\eta)^{-1}$ approaches zero. It is therefore a high-frequency approximation to the Zener model.

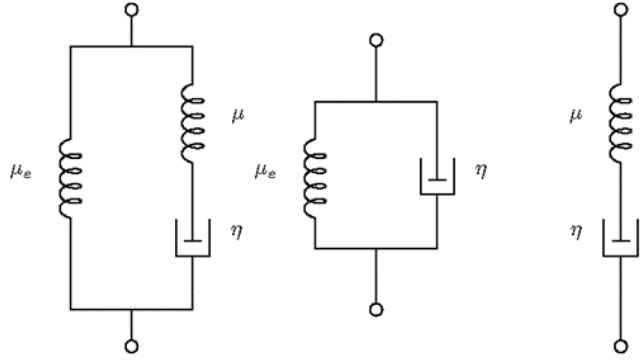

**Figure 6.** Classical models (left to right): Zener, Kelvin-Voigt, Maxwell, with constants $\mu$, $\mu_e$, $\eta$ and associated time constants $\tau_\sigma$ and $\tau_\varepsilon$.

It is common to model the elastic response of tissue with one of these elementary models. For example, the Kelvin-Voigt model (**Figure 6, middle**) has been widely used as a model, and by inspection we can write its stress-strain response in the frequency domain as the sum of the two parallel elements:

$$\mu(\omega) = \mu_e + i\eta\omega \qquad (4)$$

From this frequency response we see immediately that the low frequency response is a constant set by the spring, and the high frequency response is dominated by the viscous term times frequency to the first power. However, tissue responses, as previewed in **Figures 1 – 4**, do not fit neatly into a $\omega^0$ (constant over frequency) or an $\omega^1$ power. Thus, curve-fitting to this function invariably forces the data into the transition zone between the two regimes, around $\omega_0 = \mu_e/\eta$, however the extrapolation outside of this transition frequency is questionable. Furthermore, a major conceptual problem is that *single* relaxation time constant mechanisms are rare and found only in isolated conditions, such as certain ionic molecules in water (Ainslie and McColm, 1998; Blackstock, 2000). No one has demonstrated and identified a *single* relaxation time constant for soft tissues, in fact the opposite is true: the field of biomechanics largely abandoned this as overly simplistic many decades ago. See Fung (1981b), where he discusses continuous relaxation spectra across some extended range of relaxation time constants or frequencies and says: "I have the experience that the relaxation spectra. . . and associated properties. . .work very well in virtually all cases we know." It is time for the same approach to be adopted in elastography as well. The multiple or continuously varying relaxation models build on the reality that tissue has a hierarchy of structures which can be measured at different temporal resolving powers or equivalently, bandwidths. For each measurement, some range of structure is averaged.

*3.3 The multiple relaxation model*

The multiple relaxation approach of **Figure 7** is therefore one way to model tissue in small strains and shear wave propagation. It is a general approach that can model arbitrary frequency responses, but often it is found that the net effect of the multiple processes is a *power law* in the frequency domain. That is the link to the second approach: generalization of the viscoelastic models to fractional ones. Both approaches will be discussed here and we will show that over the limited frequency range over which one has measurements, the two approaches often cannot be distinguished from each other, however there is compelling natural evidence (see **Figure 5**) that can aid in model selection.

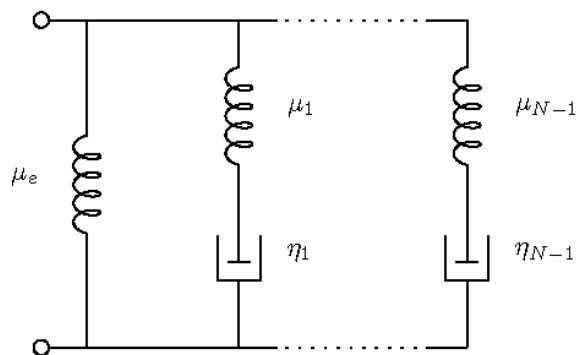

**Figure 7. Maxwell-Wiechert model.**

Many Zener models in parallel as in **Figure 7** result in a relaxation modulus which is a Prony series in the time domain with real, positive coefficients:

$$\sigma_{SR}(t) = \mu_e + \sum_{n=1}^{N-1} \mu_n e^{-t/\tau_n}, \qquad (5)$$

where $\mu_e$ is the equilibrium modulus, the asymptotic value. The dynamic modulus is also a sum:

$$\tilde{\mu}(\omega) = \mu_e + \sum_{n=1}^{N-1} \frac{\mu_n i\omega\tau_n}{1 + i\omega\tau_n}. \qquad (6)$$

This is the most general conventional model and it is called the Maxwell-Wiechert model. It is sometimes also called the generalized Maxwell or the generalized Kelvin-Voigt model and sometimes it is depicted in its conjugate form with a series combination of springs and dashpots in parallel. These models are all equivalent (Tschoegl, 1989a; Holm, 2019c). We will show in later sections that a reasonable distribution of elements in **Figure 7** and time constants $\tau_n$ leads naturally to fractional models.

*3.4 Fractional models*

The fractional models of linear viscoelasticity, the fractional Zener, fractional Kelvin-Voigt, and spring-pot models of **Figure 8** are alternative models.

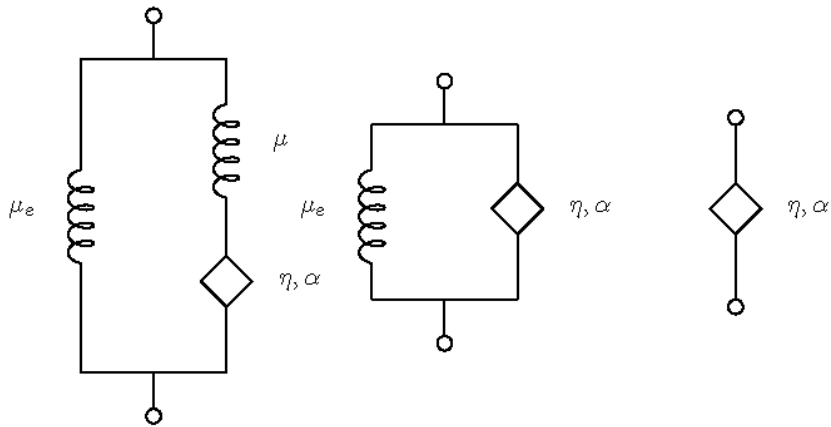

**Figure 8. Fractional Zener and Kelvin-Voigt models, and spring-pot model.**

*3.4.1 Frequency domain*

The fractional Zener model (**Figure 8, left**) is related to the standard Zener model (**Figure 6, left**) by replacing the classical damper ($i\omega\eta$ frequency response) with a fractional derivative damper ($(i\omega)^{\alpha} \cdot \eta$ frequency response where $0 < \alpha \leq 1$). It can be shown that the fractional Zener model's dynamic modulus is given by

$$\tilde{\mu}(\omega) = \mu_e \frac{1+(i\omega\tau_\varepsilon)^{\alpha}}{1+(i\omega\tau_\sigma)^{\alpha}}, \tag{7}$$

where the relationship between the two time constants of eqn (9) and the discrete components of the fractional Zener model in **Figure 8** are:

$$\tau_\sigma^{\alpha} = \frac{\eta}{\mu}, \quad \tau_\varepsilon^{\alpha} = \eta\left(\frac{1}{\mu}+\frac{1}{\mu_e}\right) \geq \tau_\sigma^{\alpha}. \tag{8}$$

This four-parameter model has some possible simplifications. If the right side spring element $(\mu)$ is assumed to be much stiffer than the other elements (over the range of frequencies studied), then the forces and displacements are substantially transferred to the viscous spring-pot element, and so the model reduces to **Figure 8, middle**, which is the KVFD model. Going further, if the left side spring element $(\mu_e)$ is assumed to be negligible compared to the viscous spring-pot, then the model reduces to **Figure 8, right**.

Correspondingly, as these approximations are made, the dynamic moduli of the three models can be simplified as follows:

$$\tilde{\mu}(\omega) = \mu_e \frac{1+(i\omega\tau_\varepsilon)^{\alpha}}{1+(i\omega\tau_\sigma)^{\alpha}} \approx \mu_e + (i\omega\eta)^{\alpha} \approx (i\omega\eta)^{\alpha}. \tag{9}$$

The first approximation is from the fractional Zener model to the fractional Kelvin-Voigt model. They both have the same low-frequency asymptote, $\tilde{\mu}(0) = \mu_e$. However the finite high-frequency asymptote of the Zener model, $\tilde{\mu}(\infty) = \mu_e (\tau_\varepsilon/\tau_\sigma)^\alpha$, is not achieved by the fractional Kelvin-Voigt model. In fact, in the literature the Kelvin-Voigt family of models has been considered to be non-standard because they have an impulse in the relaxation modulus at time 0. For this reason, Tschoegl (1989b) states that such a model "appears to be physically unrealistic." As this refers to a situation where the continuum assumption breaks down, the model may, despite this reservation, often be quite useful in practice. As an example, two parameters derived from the fractional Kelvin-Voigt model were found to be beneficial in discriminating between hepatic fibrosis and inflammation in patients with chronic liver disease in MRE (Sinkus *et al.*, 2018).

The last approximation, that of the spring-pot alone, cannot model the static, low frequency properties of the material as it is a fluid model. It is therefore a mid-frequency approximation, but quite useful over typical shear wave frequencies. For example, in two papers by Zhang *et al.* covering prostate and liver specimens (Zhang *et al.*, 2007; Zhang *et al.*, 2008), the $\mu_0$ (or $E_0$ in their notation) was found to be negligible, below 1 Pa as inferred from the mechanical stress relaxation test. Their highest recorded value was from one case of advanced prostate cancer with an $E_0$ of 700 Pa. Thus, in many situations involving normal soft tissues and frequencies within the common elastography range (40 - 1000 Hz), the two-parameter spring-pot may be sufficiently useful and justifiable.

**Figure 9** shows that the fractional Kelvin-Voigt and the fractional Zener models follow each other at low frequencies, and that the spring-pot and the fractional Kelvin-Voigt models are the same at high frequencies. But for the intermediate frequencies from about $10^0$ to $10^5$ (given by the ratio of the time constants in the fractional Zener model, which in this example is $\tau_\varepsilon / \tau_\sigma = 10^5$), and in particular $10^1 - 10^4$, all three models have more or less the same performance. This is an argument for why the theoretically correct fractional Zener model is too complex and that with only a band-limited set of measurements, the fractional Kelvin-Voigt and even the spring-pot models are adequate if one wants to use fractional models. The same performance can also be achieved with a few relaxation processes. Finally, the simpler models of **Figure 6** can be considered subsets of the fractional models. For example, the fractional Kelvin-Voigt model reduces to the Kelvin-Voigt model when $\alpha = 1$.

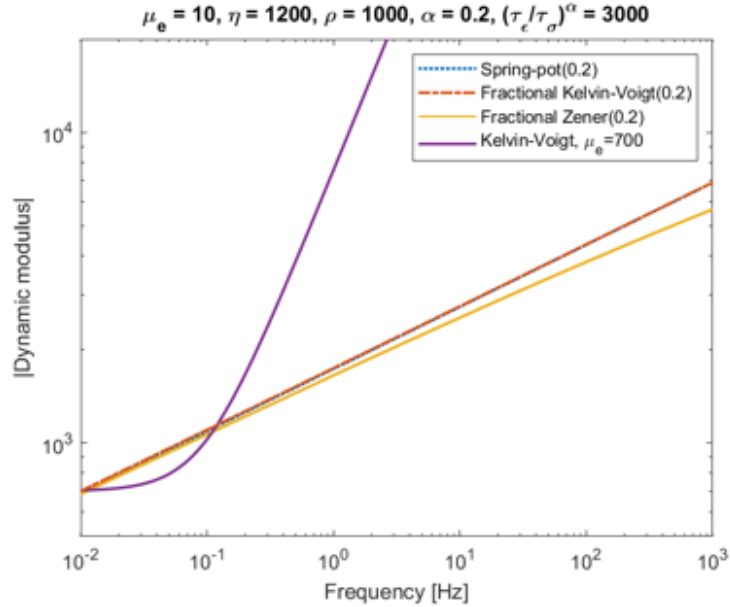

**Figure 9. Dynamic modulus for the three fractional models where $\tau = \tau_\varepsilon$ is a characteristic time constant. The spring $\mu$ is set to 10 Pa for fractional models consistent with a low value often estimated in studies. The spring is increased to 700 Pa in the linear model to demonstrate the transition and asymptotes inherent in that model.**

*3.4.2 Time domain behaviors*

The stress relaxation response of the fractional Kelvin-Voigt model is:

$$\sigma_{SR}(t) = \mu_e \left( 1 + \tau^\alpha \frac{t^{-\alpha}}{\Gamma(1-\alpha)} \right), \quad \tau^\alpha = \frac{\eta}{\mu_e}. \tag{10}$$

This is a power function with a singularity as $t$ approaches 0. This anomaly is the time domain equivalent of the infinite dynamic modulus and the infinite phase velocity as frequency increases of the Kelvin-Voigt family of models.

The fractional Zener model's relaxation modulus is given by a Mittag-Leffler function:

$$\sigma_{SR}(t) = \mu_e + \mu_e \left[\left(\frac{\tau_\varepsilon}{\tau_\sigma}\right)^\alpha - 1\right] E_\alpha\left[-(t/\tau_\sigma)^\alpha\right]. \tag{11}$$

The Mittag-Leffler function, $E_\alpha$, is a generalization of the exponential function and for $\alpha = 1$ it is the exponential function. The Mittag-Leffler function is well-behaved from the start, i.e. for all $t \geq 0$, as can be seen in **Figure 10.**

The Mittag-Leffler function with a transformed argument can be approximated:

$$e_\alpha(t) = E\alpha(-t^\alpha) \sim \begin{cases} \exp\left[\dfrac{-t^\alpha}{\Gamma(1+\alpha)}\right] \approx 1 - \dfrac{t^\alpha}{\Gamma(1+\alpha)}, & t \to 0 \\ \dfrac{t^{-\alpha}}{\Gamma(1-\alpha)}, & t \to \infty \end{cases}. \tag{12}$$

The approximation for small time is a stretched exponential, while for large values of $t$ the Mittag-Leffler function approaches a power law (Mainardi, 2014).

Combining the approximation of eqn (12) for large time with (11) gives (10), the fractional Kelvin-Voigt relaxation modulus, which is thus seen to be the large time approximation to the fractional Zener model.

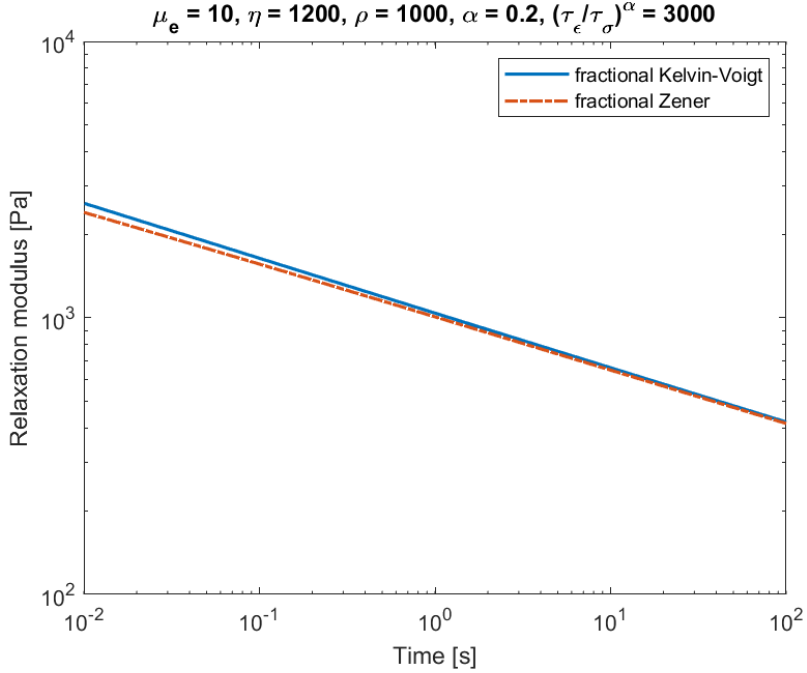

**Figure 10. Stress relaxation response (or relaxation moduli) of the classical Zener model with an exponential time response (solid line) and for the fractional Zener model for $\alpha = 0.2$ (dashed line), which asymptotically approaches a power law function.**

*3.5 Relaxation spectrum*

In the decomposition of the relaxation modulus of eqn (5) the first term characterizes the left-hand spring directly across the terminals in **Figure 7**, and the sum is for all the series connected springs and dampers.

Let the sum of eqn (5) be decomposed into

$$\sigma_{SR}(t) = \mu_e + \sigma_{SR,\tau}(t), \tag{13}$$

and decompose the time-varying part.

*3.5.1 Relaxation time-spectral function*

The continuous generalization of the multiple relaxation model is:

$$\sigma_{SR,\tau}(t) = \sigma_{SR,\tau}(0) \int_0^\infty R_\sigma(\tau) \mathbf{e}^{-t/\tau} d\tau, \tag{14}$$

where $\sigma_{SR,\tau}(0)$ is a non-negative constant and $R_\sigma(\tau)$ is a non-negative relaxation spectrum.

In (Gross, 1947; Caputo and Mainardi, 1971) as well as in (Mainardi, 2010b), it has been shown that for the fractional Zener model, the relaxation time-spectral function of eqn (14) is:

$$R_\sigma(\tau) = \frac{1}{\pi\tau} \frac{\sin\alpha\pi}{(\tau/\tau_\sigma)^\alpha + (\tau/\tau_\sigma)^{-\alpha} + 2\cos\alpha\pi} \sim \begin{cases} \dfrac{\sin\alpha\pi}{\pi\tau_\sigma^\alpha} \cdot \tau^{\alpha-1} & \text{for } \tau/\tau_\sigma \ll 1 \\ \dfrac{\sin\alpha\pi}{\pi\tau_\sigma^{-\alpha}} \cdot \tau^{-\alpha-1} & \text{for } \tau/\tau_\sigma \gg 1 \end{cases}, \quad (15)$$

and $\sigma_{SR,\tau}(0) = \mu_e\left((\tau_\varepsilon/\tau_\sigma)^\alpha - 1\right)$. The exact result of (15) inserted back into (14) yields the relaxation function of the fractional Zener model of (11).

Now consider the result for $\tau/\tau_\sigma \ll 1$, i.e. let $R_\sigma(\tau) \propto \tau^{\alpha-1}$, a power law distribution. One rationale for introducing this function is that the power law distribution is frequently found to describe multi-scale and fractal systems in biological systems (West *et al.*, 1999; Carstensen and Schwan, 1959). When inserted back into eqn (14), (including the constant $\mu_e$) this yields the relaxation function of the fractional Kelvin-Voigt model of eqn (10). Similarly, the microchannel flow model (Parker, 2014; Parker *et al.*, 2016; Parker *et al.*, 2018a) uses a continuous distribution of relaxation time constants, consistent with **Figure 7**, where the time constants are linked by Poiseuille's Law to the flow within a fractal branching vasculature and fluid channels. Under appropriate assumptions, the result resembles the fractional Kelvin-Voigt or the spring-pot, **Figures 8 middle and right**.

*3.5.2 Relaxation frequency-spectral function*

An alternative decomposition is the frequency-spectral function decomposition is

$$\sigma_{SR,\tau}(t) = \int_0^\infty S_\sigma(\Omega)\mathbf{e}^{-st} d\Omega, \qquad (16)$$

where $\Omega = 1/\tau$ represents relaxation frequencies within a distribution. It is given by the relaxation spectrum, which derives directly from the time-spectral function (Mainardi, 2010a). For the fractional Zener model of eqn (15), this can be represented as (Nasholm and Holm, 2011):

$$S_\sigma(\Omega) = \sigma_{SR,\tau}(0) \frac{R_\sigma(1/\Omega)}{\Omega^2} \sim \sigma_{SR,\tau}(0) \begin{cases} \dfrac{\sin \alpha\pi}{\pi \tau_\sigma^{-\alpha}} \cdot \Omega^{\alpha-1} & \text{for} \quad \Omega\tau_\sigma \ll 1 \\ \dfrac{\sin \alpha\pi}{\pi \tau_\sigma^\alpha} \Omega^{-\alpha-1} & \text{for} \quad \Omega\tau_\sigma \gg 1 \end{cases}. \quad (17)$$

The result may be simplified to describe a fractional Kelvin-Voigt model over a band-limited region, $\tilde{\mu}(\omega) = \mu_e + (i\omega)^\alpha$, by letting the relaxation frequencies be spread evenly over the desired frequency range on a logarithmic axis:

$$\Omega_n = 1/\tau_n = \Delta \cdot \Omega_{n-1}. \quad (18)$$

The logarithmic spread of relaxation frequencies implies that $d\Omega \propto \Omega$ in eqn (16) (Nasholm, 2013). Combining that with the asymptotic result for $\Omega\tau_\sigma \ll 1$ from (17) implies that the effective individual elastic moduli should vary according to (Holm, 2019e):

$$\mu_n \propto \Omega^\alpha. \quad (19)$$

An example with four relaxation processes ($N = 5$) is shown in **Figure 11**. Here the ratio is $\Delta = 6$ between the relaxation frequencies, from the starting value $\Omega_1 = 20$ to $\Omega_4 = 4320$. The relaxation frequencies are shown by stars in the figure. In the range from the lowest to almost the highest relaxation frequency, one cannot really distinguish the elastic modulus from that of a power law. This demonstrates that a multiple relaxation model can be just as good as a power law over such a limited frequency range.

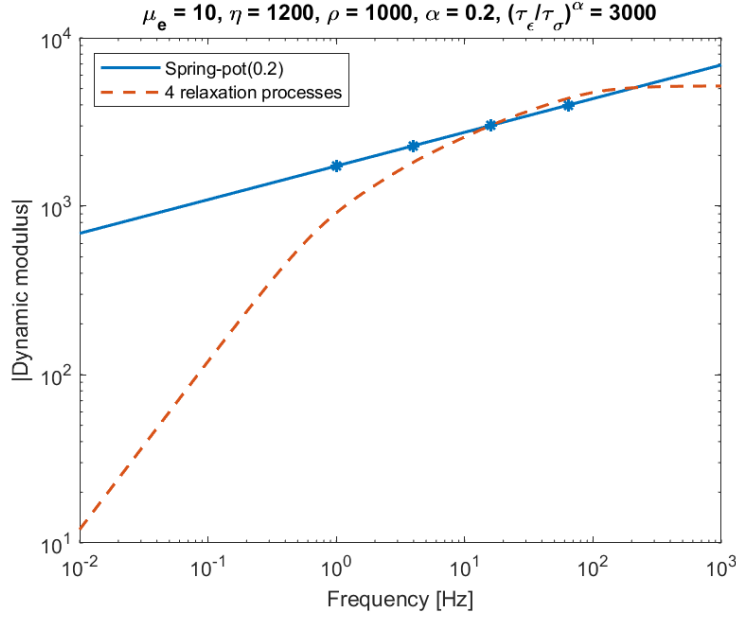

**Figure 11.** Power law dynamic response with power $\alpha = 0.2$ (fractional spring-pot) from Figure 8 and the sum of a constant + four relaxation processes where the relaxation frequencies are spread evenly in log-frequency. The characteristic time constant is $\tau = \tau_\varepsilon$.

*3.6 Fractional derivative wave models*

The complex wave number in linear wave propagation is (Holm and Holm, 2017):

$$k^2(\omega) = \frac{\rho_0 \omega^2}{\tilde{\mu}(\omega)}, \tag{20}$$

where $k$ has a real and imaginary part $k = \beta - i\alpha$ as described in more detail in **Appendix 1**. Each of the classical and multiple relaxation models considered in the previous sections prescribe a frequency dependence of the real and imaginary parts of $\tilde{\mu}(\omega)$, thus determining the complex behavior of wavenumber, $k$. From this, the real and imaginary parts of $k$ are linked to the phase velocity and attenuation coefficient, respectively (Szabo and Wu, 2000; Szabo, 2014; Holm, 2019a). The physical basis for this wave-based propagation model is attenuation being power law absorption, which, for shear waves is $\alpha_s(\omega) = \alpha_0 |\omega|^y$ in which typically $0 \leq y \leq 2$. Application of this model to data for shear and longitudinal wave absorption and dispersion for viscoelastic media was published by Szabo and Wu (2000), and earlier, for power law media (Szabo, 1994, 1995) including the cases where $0 \leq y \leq 1$. The general relations are reviewed in **Appendix 1**. A power law model which relates the elastic constant and a viscosity through causality successfully describes the observed behavior of longitudinal pressure waves in soft tissue as well as many other fluids and materials (Nachman *et al.*, 1990; Szabo, 1993, 1994, 1995; He, 1998; Szabo and Wu, 2000; Waters *et al.*, 2000; Norton and Novarini, 2003). This model was applied to magnetic resonance elastography (MRE) by Sinkus *et al.* (2007). This power law model has been shown to be equivalent to the fractional Kelvin-Voigt model with $y = \alpha + 1$ for $1 < y \leq 2$ and small values of $\omega\tau$. For large values of $\omega\tau$, the form is the same as the fractional Kelvin-Voigt model but with $y = 1 - \alpha/2$ (Holm and Sinkus, 2010; Holm, 2019d)

As in eqn (9), for a power law dynamic modulus $\tilde{\mu}(\omega) \propto \omega^{\alpha}$, the attenuation (the imaginary part of $k$) will mostly follow $\omega^{\alpha+1}$. The phase velocity on the other hand will, to a first approximation, follow a constant plus a term which is proportional to $\cos(\pi\alpha/2)\omega^{\alpha}$, meaning that there will be no dispersion in the special case of $\alpha = 1$ (Szabo, 1994; Szabo, 2014; Holm, 2019d).

## 4. Recommendations

Let us assume that a research project has carefully measured some soft tissue, small strain response over a limited range of time or frequencies. For example the soft tissue can be measured in sinusoidal steady state shear wave excitation, transient shear wave excitation, or in a stress relaxation protocol. Which rheological model out of many should the researcher consider fitting the results into a meaningful framework? Our advice is to avoid classical single relaxation models, tempting and familiar as they may be, because the support for a *single* relaxation mechanism and its characteristic behavior over reasonable frequency spans is lacking. Conversely, the historical evidence and rationale for multiple relaxation mechanisms is strong and the associated models are simple yet remarkably predictive. The ladder of complexity in this class is straightforward as indicated in eqn (9), for a two-parameter fit, the spring-pot, **Figure 8 (left)**, can be considered for large values of $\omega\tau$. For a three-parameter fit, the KVFD model, **Figure 8 (center)**, where $\mu_e$ is more significant, may be more appropriate. For a more general approach, the four-parameter fractional Zener model, **Figure 8 (left)** is well-supported, and simplifications of this to the other two are straightforward. For completion we summarize the limits for the fractional Zener model here:

- The asymptotes of the shear wave phase velocity, $c_{ph}$, are (Holm, 2019d):

$$c_{ph} \begin{cases} = c_0 \left(1 + \frac{1}{2}\cos\frac{\pi a}{2}\left[(\omega\tau_\varepsilon)^\alpha - (\omega\tau_\sigma)^\alpha\right]\right), & (\omega\tau_\varepsilon)^\alpha \ll 1 \\ \propto \omega^{\alpha/2}, & (\omega\tau_\sigma)^\alpha \ll 1 \ll (\omega\tau_\varepsilon)^\alpha . \\ = c_0 (\tau_\varepsilon/\tau_\sigma)^{\alpha/2}, & 1 \ll (\omega\tau_\sigma)^\alpha \end{cases} \qquad (21)$$

- The asymptotes of the attenuation are:

$$\alpha_k \begin{cases} \propto \omega^{1+\alpha}, & (\omega\tau_\varepsilon)^\alpha \ll 1 \\ \propto \omega^{1-\alpha/2}, & (\omega\tau_\sigma)^\alpha \ll 1 \ll (\omega\tau_\varepsilon)^\alpha , \\ \propto \omega^{1-\alpha}, & 1 \ll (\omega\tau_\sigma)^\alpha \end{cases} \qquad (22)$$

where $\alpha_k$ is used for the imaginary part of the wavenumber in order to distinguish it from the model order, $\alpha$. The upper of the asymptotic expressions, valid for $(\omega\tau_\varepsilon)^\alpha \ll 1$, may be important in practice. However, the middle range, second set of terms, are sometimes suitable for soft tissues in the common elastography range of 40 – 1000 Hz. In that case, phase velocity, group velocity, attenuation, and shear modulus are all described by only two parameters and are simply interrelated (Parker *et al.*, 2018b).

The fractional Kelvin-Voigt model has $\tau_\sigma = 0$ and therefore shares the two upper asymptotes with the fractional Zener model as the range of the middle asymptote will extend to infinite frequency.

Simplifications of these, especially where the spring elements are out of range compared to the fractional term, lead to the two simpler fractional models, the spring-pot and the fractional Kelvin-Voigt, as suitable for the range of data encountered in shear wave elastography. Compare the similarity of time domain behavior of **Figure 10** to the data of **Figure 1**. Similarly, the same model behaviors plotted in the frequency domain in **Figure 9** compare favorably with the observed behavior in **Figures 2 – 4**. As indicated in the compilation of data in **Figure 5**, which extends over four decades of frequency, the measurements are more likely to be better described by the fractional models than the multiple relaxation approach illustrated in **Figure 11**, which fits over a limited bandwidth and deviates from power law behavior elsewhere.

## 5. Discussion

### 5.1 Calculus vs. rheology

We mention for the purpose of completion that it is not always necessary to have a well-supported rheological model, particularly for very narrow bandwidth waves. From a mathematical point of view, the result in calculus known commonly as the Taylor series or Maclaurin series expansion states that any well-behaved function around a particular point can be represented by a sum that includes first order and higher order derivatives. So, in principle, the dispersion of shear wave speed, $c(\omega)$, around a particular frequency $\omega_0$, can be approximated to first order as $c(\omega) = c(\omega_0) + (\omega - \omega_0) dc/d\omega$, and similar relations hold for other parameters such as the shear modulus and attenuation. This approximation will be valid over some limited bandwidth around $\omega_0$. For weakly attenuating materials such as pure gelatin phantoms, it could be been assumed that $c$ is constant and attenuation increases linearly with frequency over some limited bandwidth. In this case the wave equation and Kramers-Kronig relations are not strictly satisfied, only as first order approximations (Blackstock, 2000).

*5.2 Previous comparisons*

While considerable effort has been expended towards obtaining repeatable measurements and determining good practices for shear wave elastography (RSNA/QIBA, 2012; Hall *et al.*, 2013; Palmeri *et al.*, 2015) there is less agreement on which models are the most applicable. Because of variety of types of shear viscoelastic measurements, especially for wave-based methods, data extraction is not as straightforward as with mechanical and testing and MRE. Our goals in this work has been to take a broader perspective, based on available data and models, and provide recommendations for viscoelastic shear measurements of soft tissues such as the liver.

Here we briefly review shear model comparisons from the elastography field. The early study by Catheline *et al*. (2004) showed that the Kelvin-Voigt model was better fit to shear wave data than the Maxwell model for a phantom and bovine muscle in the 50 to 500 Hz range. Kiss *et al*. (2004) demonstrated good agreement between their fractional Kelvin-Voigt model and normal and thermally ablated canine liver and data obtained from a dynamic mechanical testing from 0.1 to 400 Hz. Klatt *et al*. (2007) ) fit five models (Maxwell, Kelvin-Voigt, Zener, Jeffreys, and fractional Zener) to four points of MRE data (25 to 63 Hz) and slightly favored the Zener model. Sinkus *et al.*(2007) in examining breasts and breast phantoms with MRE, also with 4 data points (65 to 100 Hz), found that the relationship between the real and imaginary shear wave $G$ constants strongly supported a wave power law model not a Voigt model. Bercoff *et al.* (2004) presented a theory for shear wave generation accounting for diffraction induced by the acoustic radiation force method; therefore, they indicated that correction of attenuation data was necessary before utilizing a rheological model. Urban and Greenleaf (2009) applied a radial diffraction correction to their muscle fiber data (50 to 600 Hz) and found the power law model worked reasonably well and was similar to the Kelvin-Voigt model over the range of parameters for their experiments. Kumar *et al.* (2010) in their study of Maxwell, Kelvin-Voigt, and fractional Kelvin-Voigt models applied to measurements of Young's modulus from polyacrylamide-based phantoms found that only the fractional KV model matched their data. Urban *et al.* (2011) tested five models (Maxwell, generalized Maxwell, Zener, Kelvin-Voigt, and fractional Kelvin-Voigt) and a finite element program against time domain data for a sphere embedded in viscoelastic phantoms; however, the differences among the models for this time domain experiment were slight. Wex et al. studied preservation times of porcine liver and obtained good agreement between the fractional KV model and rheometer data from 0.1 to 10 Hz. Rouze *et al.* (2018) compared their shear wave measurements from viscoelastic phantoms, corrected for diffraction and other effects, to a power law model and found the absorption exponent to be close to 1 $\left( y \sim 1 \right)$ inconsistent with the Kelvin-Voigt model previously used (Rouze *et al.*, 2016). Finally, Parker, et al. (2018b) in a follow–on shear wave study (see **Figure 2**) on beef liver using a variety of methods (from 40 to 380 Hz) found best agreement with the high frequency approximation of the fractional Kelvin-Voigt model with a power law of approximately 0.2.

*5.3 Sensitivity of parameters*

In fitting a model to data, there are two categories that influence confidence intervals. The first is the noise or errors in the basic measurements. Errors in estimates of the shear wave amplitudes, arrival times, peaks, and related measures will propagate into uncertainties in parameter fitting (Urban *et al.*, 2009; Elegbe and McAleavey, 2013).

A second important category is the number of parameters that are being simultaneously estimated. It is generally true that there are many models for which more than two parameters are calculated simultaneously. In some of these problems the model equations are linear with respect to the parameters and for these the calculation procedure is available in any number of statistics texts (Draper and Smith, 1966) and is known as multiple linear regression. More generally, the joint confidence region for all the parameters becomes an ellipsoid in $P$ dimensional space for $P$ parameters (Scheffé, 1959). Because the confidence intervals of the $P$ unknown parameters are dependent (Schwartz, 1980), one can be more confident with fewer parameters. This matches with the general sense of Occam's razor, favoring the simplest explanation for a phenomenon. In the case of rheological models of tissue, this favors the simplest models of **Figure 8**, the fractional Kelvin-Voigt, and in some cases, the fractional spring-pot.

*5.4 Phantoms*

We make no specific recommendations for rheological models of tissue-mimicking phantoms, except to note that the composition of a phantom, its components and morphology and the manufacturing or "curing" steps employed, will all play a role in determining the complex modulus. For some phantoms (Sinkus *et al.*, 2007; Kumar *et al.*, 2010; Coussot *et al.*, 2009) and for the QIBA phantom studies (Rouze *et al.*, 2018), the fractional Kelvin-Voigt model is suitable. Given all these factors, it is difficult to make any generalizations about phantoms or to prefer those models shown in **Figures 7**, **8**, or **9**. In fact, the oil-in-gelatin models of steatosis (fat accumulation in the liver) that were analyzed by Parker *et al*. (2018a) were found to have a reasonable fit to the theory of Christensen (1969) for composite materials. He applied a principle of minimum strain energy in a deformed elastic medium with specifically spherical inclusions. This work found simplifications for the effective shear modulus in the limiting case of the volume fraction of spheres being small or large; the asymptotic approach to volume fraction of zero or one. A more recent overview of different types of composites, inclusion shapes, and results are given in Chapter 9 of Lakes (Lakes, 1999c). Composite materials are one class of possible types that may require rheological models beyond those recommended in Section 4 for normal soft tissues. Case-by-case assessment is likely required and this subject is beyond the scope of this paper.

## 6. Conclusion

Nearly 2500 years ago, the Greek philosopher Heraclitus of Ephesus proposed a universal model of rheology: πάντα ῥεῖ (*panta rhei*) "everything flows" (Beris and Giacomin, 2014) and, in fact, soft tissues under shear stress will deform in characteristic ways. We argue that single relaxation models, simple as they may be and familiar from our first exposure to stress and strain lectures, are not appropriate for modeling the response of soft tissues to small strains in the common range of elastography techniques. Instead, multiple relaxation models are recommended. These can take numerous equivalent forms in the time domain or frequency domain, from discrete series representing several parallel elements, to power law behaviors with as few as two parameters. In particular, the fractional Zener model is compact and meaningful, in the sense that its parameters can be linked to mathematical steps that summarize a superposition of relaxation mechanisms from a power law distribution, one of the most common behaviors observed in the natural world (Newman, 2005) across many phenomena. For soft tissue, the multi-scale nature of the tissue, extending from molecular chains to cellular walls to fluid channels, connective tissues, and larger structures such as arterial walls, likely combine to form the extended relaxation spectra. However, further research is needed to isolate and quantify the contributions in specific tissues.

The question of tissues in pathological states also requires further investigation at this time; we make no assertion that the fractional Zener and KVFD models will hold for all tissues affected by advanced diseases or for anisotropic tissues such as muscles and tendons. This is another rich area needing further study, motivated by the useful empirical changes in shear wave speed with pathology that have been noted in elastography studies (Barr *et al.*, 2015).

**Acknowledgements**

Kevin Parker was supported by NIH grant R21EB025290. Thomas Szabo appreciates the support of the department of Biomedical Engineering at Boston University. Sverre Holm was supported by the European Union's Horizon 2020 Research and Innovation program, Grant/Award Number: 668039. This paper reflects only the authors' view. The European Commission is not responsible for any use that may be made of the information it contains.

**APPENDIX 1: Shear wave speed and attenuation**

The simplest way to view the interrelationship between rheological models and shear wave speed and attenuation is by considering one-dimensional shear wave propagation in a linear, isotropic, homogeneous, infinite medium. In summary, using the complex exponential to represent a plane wave at frequency $\omega$, the solution to the lossless wave equations for a disturbance traveling in the $+x$ direction is

$$u(x) = A\mathbf{e}^{-i(kx-\omega t)}, \tag{23}$$

where $u(x)$ is displacement, $A$ is an amplitude, $i$ is the imaginary unit, and $k$ is the wavenumber. Furthermore,

$$k = \frac{\omega}{c} \tag{24}$$

and

$$c_s = \sqrt{\frac{\mu}{\rho}} \cong \sqrt{\frac{E_0}{3\rho}}, \tag{25}$$

where $c_s$ is the wave speed, $\rho$ is the density, $\mu$ is the shear modulus, and $E_0$ is the Young's modulus, the approximation $\mu = E_0/3$ valid for nearly incompressible materials. Additional details can be seen in Graff (1975b), or summarized for elastography in Parker *et al.* (2011). The important point is that in this purely elastic, lossless propagation, there is only one velocity $c$ for all frequencies and no distinction in speeds observed between monochromatic narrowband and broadband shear wave disturbances.

The situation changes when some loss mechanism enters the wave equation, usually through the constitutive equations of the material that is supporting the shear waves. In this case, the speed $c_s$ changes with frequency (a phenomenon called "dispersion") and the concept of group velocity is introduced, related to the slope, or derivative, of phase velocity $c_p$. An excellent treatment is given in section 1.6.1 of Graff (1975a).

Lossy or viscoelastic materials like tissues can be characterized by their stress-strain relations leading to complex $E(\omega)$ and $\mu(\omega)$ (Lakes, 1999b). In the viscoelastic case, the solution to the lossy wave equation still resembles eqn (23), but now $k$ is complex: its real component is still related to $\omega/c_p$ and its imaginary component defines an exponential decay with distance. Accordingly, we write a general form for complex, frequency-dependent shear modulus (Lakes, 1999b; Zhang and Holm, 2016):

$$\mu^*(\omega) = \left(\mu_d(\omega) + i\mu_i(\omega)\right), \tag{26}$$

where $\mu^*$ is the complex modulus, $\mu_d$ is the dynamic modulus, and $\mu_i$ is the loss modulus. So the propagation constant

$$k = \frac{\omega}{\sqrt{\frac{\mu_d(\omega) + j\mu_i(\omega)}{\rho}}} = \beta - i\alpha = \frac{\omega}{c_p} - i\alpha, \tag{27}$$

and the wave now propagates as

$$u(x) = A\mathbf{e}^{-\alpha x}\mathbf{e}^{i(\omega t - \beta x)}$$
$$= A\mathbf{e}^{-\alpha x}\mathbf{e}^{i\omega\left(t - \frac{x}{\omega/\beta}\right)}. \tag{28}$$

The latter form emphasizes that this is a rightward $(+x)$ propagating wave with speed of $c_p = \omega/\beta$. Now sorting through the real and imaginary parts of eqn (27) and denoting

$$|\mu| = \sqrt{\mu_d^2(\omega) + \mu_i^2(\omega)}, \tag{29}$$

we find that:

$$\beta = \left[\omega\sqrt{\frac{\rho}{|\mu|}}\right]\left[\frac{1}{2}\left(1+\frac{\mu_d(\omega)}{|\mu|}\right)\right]^{1/2}. \qquad (30)$$

The phase velocity is a function of frequency

$$c_p = \left[\sqrt{\frac{|\mu|}{\rho}}\right]\left[\frac{1}{2}\left(1+\frac{\mu_d(\omega)}{|\mu|}\right)\right]^{-1/2}, \qquad (31)$$

and the attenuation coefficient has a leading term directly proportional to the first power of frequency (even though more complicated dependencies on frequency remain hidden in $\mu_d$):

$$\alpha = \left[\omega\sqrt{\frac{\rho}{|\mu|}}\right]\left[\frac{1}{2}\left(1-\frac{\mu_d(\omega)}{|\mu|}\right)\right]^{1/2}. \qquad (32)$$

Particular forms of these for some tissue models can be found in Carstensen and Parker (2014), and these simplify for power law behaviors (see eqn (21) and (22), Parker *et al.* (2018b), and Holm (2019d)). Equations (9), (31), (32) provide a bridge of equivalency between the mechanical and wave parameters.

In more general terms, the constraints of causality impose specific inter-relationships between attenuation and phase velocity as a function of frequency. These can be captured in the form of Kramers-Kronig relations (Szabo, 1993, 1994, 1995; Szabo and Wu, 2000; Waters *et al.*, 2000; Cobbold, 2007), and through Hilbert transform relations (Bracewell, 1965).

**APPENDIX 2: Glossary**

| Symbol | Name | Comments |
| --- | --- | --- |
| $E_0$ | Young's modulus | Standard linear elastic solid stiffness, related to shear wave speed (squared). |
| $E(\omega)$ | Complex modulus | Viscoelastic and frequency-dependent stiffness measured by sinusoidal excitation in rheology; imaginary term related to loss, attenuation. Sometimes called dynamic modulus. |
| $E'(\omega)$ | Storage modulus | Real part of $E(\omega)$. |
| $E''(\omega)$ | Loss modulus | Imaginary part of $E(\omega)$. |
| $\mu(\omega), G(\omega)$ | Complex shear modulus | Related to shear wave speed (squared); real and imaginary parts termed "storage" and "loss" moduli, respectively. Approaches $E(\omega)/3$ for an incompressible elastic media. |
| $c_s$ | Shear wave speed | In an elastic medium |
| $c_p$ | Shear wave phase velocity | Related to $\mu(\omega)$ and $E(\omega)$ by eqn (31). |
| $\alpha_k$ | Shear wave attenuation | Loss, related to $\mu(\omega)$ and $E(\omega)$ by eqn (32). |
| $\beta_k$ | Shear wave dispersion | Dispersion related to the real parts of $\mu(\omega)$ and $E(\omega)$. |
| $\sigma_{SR}(t)$ | Stress relaxation vs. time | Standard step response test; can be Fourier-transformed and curve-fit to estimate $E(\omega)$ and $\mu(\omega)$. |
| $\sigma_{IR}(t)$ | Impulse response of tissue in shear | Idealized linear response; the time derivative of the step response. Its Fourier transform is $\mu(\omega)$. |
| $\tilde{\sigma}_{SR}(\omega)$ | Fourier transform of $\sigma_{SR}(t)$ | Stress relaxation modulus. |
| $\tilde{\sigma}_{IR}(\omega)$ | $i\omega \cdot \tilde{\sigma}_{SR}(\omega) = \tilde{\mu}(\omega)$ | Impulse response is the derivative of step response. |
| $i\omega$ | Fourier transform of $d/dt$ operator | First temporal derivative operator. |
| $\alpha$ | Fractional exponent | Exponent used in fractional Kelvin-Voigt and Zener models. |
| $(i\omega)^\alpha$ | Fourier transform of fractional derivative | important in fractional derivative and power law models; can be derived from multiple relaxation models |
| $y$ | Power law exponent | Exponent used in power law models. |
| $\eta$ | Viscosity | Viscosity parameter used in viscoelastic models. |
| $\tau$ | Relaxation constant | Relaxation parameter used in fractional Kelvin-Voigt, Zener and multiple relaxation models ($\tau = \eta/\mu$). |
| $\rho$ | Density | Density used in wave equations and models. |